\documentclass[nofootinbib,preprintnumbers,amsmath,amssymb]{revtex4}

\usepackage{float}
\usepackage{graphicx}
\usepackage{amsmath}
\usepackage{amssymb}
\usepackage{epsf,latexsym}
\usepackage{times}
\usepackage{bm}

\usepackage{float}

\newcommand{\ket}[1]{\left| #1 \right\rangle}
\newcommand{\bra}[1]{\left\langle #1 \right|}

\newcommand{\be}{\begin{equation}}
\newcommand{\ee}{\end{equation}}
\newcommand{\ba}{\begin{eqnarray}}
\newcommand{\ea}{\end{eqnarray}}

\begin{document}
\title{The entanglement of few-particle systems when using the local-density approximation}

\author{J P Coe} 
\email{jpc503@york.ac.uk}

\author{I D'Amico}
\email{ida500@york.ac.uk}
\affiliation{Department of Physics, University of York, York YO10 5DD, United Kingdom.}

\begin{abstract}
In this chapter we discuss methods  to calculate the entanglement of a system using density-functional theory. We firstly introduce density-functional theory and the local-density approximation (LDA). 
We then discuss the concept of the `interacting LDA system'. This is characterised by an interacting many-body Hamiltonian which reproduces, uniquely and exactly, the ground state density obtained from the single-particle  Kohn-Sham equations of density-functional theory when the local-density approximation is used.  We motivate why this idea can be useful for appraising the local-density approximation  in many-body physics particularly with regards to entanglement and related quantum information applications.  Using an iterative scheme, we find the Hamiltonian characterising the interacting LDA system in relation to the test systems of Hooke's atom and helium-like atoms.  The interacting LDA system  ground state wavefunction is then used to calculate the spatial entanglement and the results are compared and contrasted with the exact entanglement for the two test systems.  For Hooke's atom we also compare the entanglement to our previous estimates of an LDA entanglement.  These were obtained using a combination of evolutionary algorithm and gradient descent, and using an LDA-based perturbative approach.  We finally discuss if the position-space information entropy of the density---which can be obtained directly from the system density and hence easily from density-functional theory methods---can be considered as a proxy measure for the spatial entanglement for the test systems.
\end{abstract}
\maketitle
\section{Introduction}
 The mapping of an interacting many-body system to a collection of effective single particle equations is a ubiquitous and powerful approach in theoretical physics.  This is perhaps most notably embodied in density-functional theory (DFT) where a non-interacting many-body system, the Kohn-Sham system \cite{KS}, is proven to have exactly the same ground state density as the original interacting system.  By solving a set of single-particle Kohn-Sham (KS) equations, the ground state density can be calculated and from this in principle all the ground-state properties of the interacting many-body system \cite{HK}.  The KS equations would give the exact density due to the inclusion in their effective single-particle potential of the so-called  ``exchange-correlation'' potential ($v_{\text{xc}}$).  In practice $v_{\text{xc}}$ is an unknown functional of the density so approximations for $v_{\text{xc}}$ have been designed. The most popular and often effective method is the local-density approximation (LDA).  
DFT has been so far one of the most powerful methods to calculate many-body properties of complex systems, and its efficiency
lies in the observation that modelling many-particle systems exactly is generally computationally intractable as, for example, one is dealing with a many-body wavefunction whose storage requires excessive amounts of memory.  Solving a system of single-particle equations or dealing with the density, which is a function of only one variable, represents indeed a computationally attractive alternative. However we do not know in general how to express some ground state properties, such as the entanglement, in terms of the ground state particle density, and this implies that common approximation schemes used within DFT do not have a straightforward extension to the calculations of these properties. 
Here we present a scheme to calculate the entanglement associated with the LDA approximation and
investigate how well it compares with the exact entanglement.  Our test-bed systems are helium-like atoms and Hooke's atoms.   This information could be useful when considering which approximation to invoke to model the suitability of systems as a component of a quantum information processing device.  As the entanglement is considered as one of the main resources for these devices, the knowledge of how much entanglement could be created in a certain system appears essential for quantum information purposes.  In addition comparison of the entanglement associated with a DFT approximation to the exact entanglement offers another way to appraise the accuracy of the approximation.    

In this chapter we firstly introduce DFT and the LDA.  We explain the concept of the `interacting LDA system' (i-LDA) and how this may be used to calculate the entanglement corresponding to the use of the LDA.  We then introduce our test-bed systems,   Hooke's atom  and the helium-like atoms.  Next we consider how these systems can be solved using numerically `exact' methods and also approximately using the LDA within DFT. The exact and i-LDA spatial entanglement results are then displayed and compared.  We contrast these results with the position space information entropy which can easily be expressed as a functional of the density---and hence calculated using traditional DFT schemes---and investigate its usefulness as a proxy measure for the entanglement.  Finally, we compare the i-LDA entanglement to our previous best approximations to an LDA entanglement, found using a combination of evolutionary algorithm and gradient descent, and an LDA-based perturbative approach \cite{Coe}.

\section{Density-functional theory}
\label{sec:chap1DFT}
Directly solving the many-body Schr\"{o}dinger equation is a formidable, arduous task and doing so accurately for systems of more than a few electrons remains outside the realms of contemporary computing.  This can be partly illuminated by considering the wavefunction on a mesh of $1000^{3}$ points for one electron; for the same accuracy with $N$ electrons one would require $1000^{3N}$ points, clearly an excessive amount of memory as $N$ becomes large.  Yet nature, in a sense performs similar calculations all the time, but these are too complex to be reproduced by `classical' computation.  Quantum computing offers some hope in the simulation of quantum systems  (see for example Ref.~\cite{Zalka98} and Ref.~\cite{NIELSEN}).  However, unless quantum computers of a substantial number of qubits are realised, simulating  large systems may not be feasible.  There is though a powerful and important method that allows classical computing to efficiently model many-body electron systems through the density: density-functional theory.  The density depends upon only one co-ordinate (so only $1000^{3}$ mesh points in our example) therefore by using this, rather than the many-body wavefunction, solving many-body systems can be made computationally more tractable.  In its simplest formulation density-functional theory expresses observables of the ground-state of an interacting electron system as functionals of the ground-state density,
\begin{equation}
n(\bm{r_{1}})=N\int|\Psi(\bm{r_{1}}...\bm{r_{N}})|^{2}\bm{dr}_{2}...\bm{dr}_{N}
\end{equation}
and is exact in principle due to a theorem by Hohenberg and Kohn \cite{HK} which we present next.  
\subsection{The Hohenberg-Kohn Theorem}

In an electron system with Hamiltonian
\begin{equation}
\hat{H}=\hat{T}+\hat{V}_{ee}+\sum_{i}v_{ext}(\bm{r}_{i})
\end{equation}
with a non-degenerate ground state, the ground state density uniquely determines the ground state many-body wavefunction and therefore the external potential.  This is known as the Hohenberg-Kohn theorem \cite{HK}.  The usual proof shows how the density uniquely determines the wavefunction \cite{DFTREVIEW}. 
Assume there exist two Hamiltonians $H_{1}$ and $H_{2}$ with different ground-state wavefunctions, $\ket{\Psi_{1}}$ and $\ket{\Psi_{2}}$ respectively, that give rise to the same density. Then

\begin{equation}
\nonumber
\bra{\Psi_{1}}H_{2}\ket{\Psi_{1}}=\bra{\Psi_{1}}H_{2}-H_{1}\ket{\Psi_{1}}+\bra{\Psi_{1}}H_{1}\ket{\Psi_{1}}
\end{equation}

\begin{equation}
\nonumber
=\bra{\Psi_{1}}\sum_{i}\left(v_{2,ext}(\bm{r}_{i})-v_{1,ext}(\bm{r}_{i})\right)\ket{\Psi_{1}}+E_{1}
\end{equation}

\begin{equation}
=\int\left( v_{2,ext}(\bm{r})-v_{1,ext}(\bm{r})\right) n(\bm{r})\bm{dr}+E_{1}.
\end{equation}

Then by the variational principle

\begin{equation}
\int\left( v_{2,ext}(\bm{r})-v_{1,ext}(\bm{r})\right)n(\bm{r})\bm{dr}+E_{1}>E_{2},
\end{equation}
with $E_2$ the ground state related to $H_{2}$.

This can be repeated but with $1$ and $2$ exchanged to give

\begin{equation}
\nonumber
\bra{\Psi_{2}}H_{1}\ket{\Psi_{2}}=\bra{\Psi_{2}}H_{1}-H_{2}\ket{\Psi_{2}}+\bra{\Psi_{2}}H_{2}\ket{\Psi_{2}}
\end{equation}

\begin{equation}
\nonumber
=\bra{\Psi_{2}}\sum_{i}\left(v_{1,ext}(\bm{r}_{i})-v_{2,ext}(\bm{r}_{i})\right)\ket{\Psi_{2}}+E_{2}
\end{equation}

\begin{equation}
=\int\left( v_{1,ext}(\bm{r})-v_{2,ext}(\bm{r})\right)n(\bm{r})\bm{dr}+E_{2}.
\end{equation}

By the variational principle

\begin{equation}
\int\left( v_{1,ext}(\bm{r})-v_{2,ext}(\bm{r})\right)n(\bm{r})\bm{dr}+E_{2}>E_{1}.
\end{equation}

Adding the inequalities gives

\begin{equation}
E_{1}+E_{2}>E_{1}+E_{2}.
\end{equation}
This inequality is a fallacy so the initial assumption must have been incorrect: there do not exist two ground state wavefunctions that give rise to the same density. Hence by reductio ad absurdum the density uniquely determines the wavefunction. 

 As we may express the external potential as

\begin{equation}
-\frac{\hat{T}\psi}{\psi}-\frac{\hat{V}_{ee}\psi}{\psi}+E=\sum_{i}v_{ext}(\bm{r}_{i})
\end{equation}
then the density by determining the wavefunction also determines the external potential (up to an additive constant), and therefore all of the properties of the system.

  Although the Hohenberg-Kohn theorem proves the existence of the one-to-one relationship between ground state densities and wavefunctions it does not suggest how to construct the density nor find the ground-state properties of the system from the density.  To achieve this the Kohn-Sham equations \cite{KS} are used.

\subsection{The Kohn-Sham equations}
\label{sec:Chap1KSequations}
The Kohn-Sham (KS) equations allow the ground-state density to be found efficiently and the system energy to be expressed using the density.   
The KS equations are derived by considering the system energy as a functional of the density $E[n]$, this is minimised subject to a constraint incorporating $N$, the number of electrons,

\begin{equation}
\int n(\bm{r})\bm{dr}=N.
\end{equation}

Using Lagrange multipliers with the functional derivative 

\begin{equation}
\frac{\delta F[n(y)]}{\delta n(x)}=\lim_{\epsilon \rightarrow 0} \frac{F[n(y)+\epsilon\delta(y-x)]-F[n(y)]}{\epsilon}
\end{equation}

gives

\begin{equation}
\frac{\delta E[n]}{\delta n(\bm{r})}-\mu=0.
\end{equation}
As the expression of the kinetic energy as a functional of the density for an interacting electron system is unknown, Kohn and Sham re-wrote the exact energy functional in terms of the kinetic energy of non-interacting electrons $T_{NI}[n]$, the Hartree energy $U[n]$, and the potential energy from the electron density.  This requires the introduction of a functional correctional term, the exchange-correlation energy $E_{xc}[n]$. The expression for $E[n]$ then becomes

\begin{equation}
E[n]=T_{NI}[n]+U[n]+\int v_{ext}(\bm{r})n(\bm{r})\bm{dr}+E_{xc}[n]
\label{eq:chp1Enonint}
\end{equation} 
where
\begin{equation}
U[n]=\frac{1}{2}\int\int \frac{n(\bm{r})n(\bm{r'})}{|\bm{r}-\bm{r'}|}\bm{dr}\bm{dr'}.
\label{eq:chp1U}
\end{equation}

Using Lagrange multipliers now gives
\begin{equation}
\frac{\delta T_{NI}[n]}{\delta n(\bm{r})}+v_{eff}(\bm{r})-\mu=0.
\label{eqn:KSlagrangeresult}
\end{equation}

Here
\begin{equation}
v_{eff}(\bm{r})=v_{\text{H}}(\bm{r};[n])+ v_{ext}(\bm{r})+v_{\text{xc}}(\bm{r};[n])
\end{equation}
where they define the exchange-correlation potential
\begin{equation}
v_{\text{xc}}(\bm{r};[n])=\frac{\delta E_{xc}[n]}{\delta n(\bm{r})}
\label{eq:Chap1vxc}
\end{equation}
and the Hartree potential 
\begin{equation}
v_{\text{H}}(\bm{r};[n])=\int \frac{n(\bm{r'})}{|\bm{r}-\bm{r'}|}\bm{dr'}.
\label{eq:vHartree}
\end{equation}

Now Eq.~\ref{eqn:KSlagrangeresult} can be read as a non-interacting system---the KS system---with a potential $v_{eff}$.  So to find the density and hence the energy the KS equations must be solved self-consistently:

\begin{equation}
\left(-\frac{1}{2}\nabla^{2}+v_{eff}(\bm{r})\right)\phi_{i}(\bm{r})=\epsilon_{i}\phi_{i}(\bm{r})
\label{eq:KS1}
\end{equation}

and
\begin{equation}
n(\bm{r})=\sum_{i=1}^{N}|\phi_{i}(\bm{r})|^{2}.
\end{equation}

As 
\begin{equation}
T_{NI}[n]=\sum_{i=1}^{N}\int \phi_{i}^{*}(\bm{r})\left(-\frac{1}{2}\nabla^{2}\phi_{i}(\bm{r})\right) \bm{dr}
\end{equation}
then the energy $E[n]$ may be calculated from the density using $E_{xc}$.

This scheme would be exact but the general functional form of $E_{xc}$ is unknown hence it has to be approximated.  One of the simplest yet often effective approximations is using the fact that, for systems with slowly varying density, the exchange-correlation density can be locally approximated by that of a uniform electron gas. This is known as the local density approximation.   

\subsection{The local-density approximation}
\label{sec:Chpt1LDA}
The local-density approximation for the exchange-correlation energy depends only upon the density as opposed to generalised gradient approximations which include first and higher derivatives. It may be written using the form
\begin{equation}
E_{xc}=\int \bm{dr} e_{xc}(n(\bm{r})).
\end{equation}
The exchange-correlation energy density may be separated into its exchange and correlation parts, $e_{xc}=e_{x}+e_{c}$. Then, for a three-dimensional homogeneous electron gas, $e_{x}$  is known exactly \cite{WIGNER} 
\begin{equation}
e_{x}=-\frac{3}{4}\left(\frac{3}{\pi}\right)^{1/3}n^{\frac{4}{3}},
\label{eq:WignerSeitz}
\end{equation}
 while $e_{c}$ has to be approximated. We use $e_{c}$ as calculated numerically
by Perdew-Wang \cite{PERDEW92} by fitting the parameters $A$, $\alpha$, $\beta_1$\ldots$\beta_4$ to the results of Monte-Carlo simulations of the homogeneous electron gas \cite{CEPALDER80}. For zero spin polarisation this has the form
\begin{equation}
e_{c}=-2A n(1+\alpha r_{s})\log\left(1+\frac{1}{2A(\beta_{1}r_{s}^{1/2}+\beta_{2}r_{s}+\beta_{3}r_{s}^{3/2}+\beta_{4}r_{s}^{2})} \right),
\label{eq:PerdewWang}
\end{equation}
where $r_{s}$ is the radius of the sphere whose volume is equal to the volume per electron in a homogeneous electron gas,
\begin{equation}
r_{s}=\left(\frac{3}{4\pi n(r)}\right)^{\frac{1}{3}}.
\end{equation}
The above approximation is exact for the homogeneous electron gas so would be expected to be more accurate for systems close to this limit.

\section{The interacting LDA system}

 We wish to compare the entanglement calculated using a local-density approximation to the exact entanglement, and although in theory the density determines all the ground-state properties of the system it is unknown how to use the density to represent certain observables, including the entanglement.  One method to calculate the entanglement within LDA is to construct that many-body interacting system whose ground state wavefunction reproduces exactly the LDA density. We named this `the interacting LDA system' (i-LDA) \cite{Coe}.  By the Hohenberg-Kohn theorem \cite{HK} this system is unique. 

Given a test-bed interacting many-body system for which ground-state properties, such as the entanglement, can be calculated exactly, by comparing their values with the ones obtained using the corresponding  i-LDA wavefunction, we can infer how precise the LDA can be as an approximation to those quantities. Notice that the external potentials characterising the initial many-body system and the i-LDA system are different, and that {\it the LDA density is at the same time an approximation to the density of the test-bed system and the exact ground state density of the i-LDA system}.   

Such a route to assess system properties within the LDA is computationally intensive, but, for small test systems, allows a way to investigate the accuracy of the LDA for quantities for which functional expressions in terms of the density are currently unknown.  In addition this procedure offers a novel way to appraise the LDA  by considering the differences between the exact and i-LDA external potentials, thereby furnishing researchers with spatially resolved information about an approximation's accuracy. A similar procedure to the one described can be applied to other approximations within DFT. 

We introduced a method to approximate the i-LDA wavefunction in Ref.~\cite{Coe}, then in Ref.~\cite{Coe2} we developed an iterative scheme to find the Hamiltonian of the i-LDA system `exactly' (within numerical accuracy). Then the `exact' i-LDA wavefunction could then be derived.  This is the scheme we use in this chapter to allow the entanglement corresponding to the LDA to be ascertained.

\section{The test-bed systems}
The helium atom comprises two electrons held by a nuclear charge arising from two protons in the nucleus.  By using the Born-Oppenheimer approximation the motion of the nucleus may be ignored.  The system generalises to helium-like atoms by considering a nuclear charge of $Z$ giving rise to the following non-relativistic Hamiltonian in atomic units  $\left (e=\hbar=m=4\pi\epsilon_{0}=1 \right)$
\begin{equation}
H=-\frac{1}{2}\nabla^{2}_{1}-\frac{1}{2}\nabla^{2}_{2}-\frac{Z}{r_{1}}-\frac{Z}{r_{2}}+\frac{1}{\left|\bm{r_{1}}-\bm{r_{2}}\right|},
\end{equation}
where $r=|\bm{r}|$. With $Z=2$ this is the Hamiltonian for the helium atom.   Interestingly although the hydride atom ($Z=1$) exists, and its energy may be calculated, Baker et al \cite{BAKER} showed that for nuclear charges below $Z\approx 0.911$ the system is unbound.

Hooke's atom is an interacting system of two electrons held by a harmonic confining potential with Hamiltonian in atomic units of
\begin{equation}
H=-\frac{1}{2}\nabla^{2}_{1}-\frac{1}{2}\nabla^{2}_{2}+\frac{1}{2}\omega^{2}r_{1}^2+\frac{1}{2}\omega^{2}r_{2}^2+\frac{1}{\left|\bm{r_{1}}-\bm{r_{2}}\right|}.
\end{equation}
Here $\omega$ is the characteristic frequency of the confining potential.  The exact spatial entanglement and its approximations within LDA  have previously been evaluated for Hooke's atom in Ref.~\cite{Coe}.

\section{Method of solution}

We solve the  many-body Schr\"{o}dinger equation, using `exact' diagonalisation (EXD) with an efficient choice of basis.  For two electrons in a spherically symmetric potential the ground state can only be a function of the distance of each electron from the origin and the angle between the electron vectors.  Hence we may find the ground state using exact diagonalisation with a basis in three dimensions: $r_{1},r_{2},\theta$.

We use the basis
\begin{equation}
\phi_{ijl}=R_{i}(r_{1})R_{j}(r_{2})\frac{1}{\sqrt{4\pi}}\sqrt{\frac{2l+1}{4\pi}}P_{l}(\cos(\theta)),\label{basis}
\end{equation}
where we employ the hydrogen-like wavefunction 
$R_{i}(r)=Q_{i}(r)e^{-\alpha r}$.
The choice of $\alpha$ is influenced by the strength of the nuclear charge $Z$.  Here the $Q_{i}(r)$ are polynomials of degree $i$ created via the Gram-Schmidt procedure such that the $R_{i}$ are orthonormal, i.e.,  
$\int_{0}^{\infty}R_{i}(r)R_{j}(r)r^{2}dr=\delta_{ij}$.  The Legendre polynomials $P_{l}$ are orthogonal to each other and we explicitly include the normalisation factor for integration over all angles.  
We intend to calculate the matrix elements of the Hamiltonian with respect to the $\phi_{ijl}$ and $\phi_{i'j'l'}$ basis functions Eq.~\ref{basis}.
The external potential is spherically symmetric therefore the potential integrals become just $1$D integrals and are only non-zero when $l=l'$. This is propitious as they are the only integrals that have to be re-calculated each time in the iterative scheme.

The kinetic term in the integrand is
\begin{equation}
\nabla^{2}_{i}=\frac{1}{r^{2}_{i}}\frac{\partial}{\partial r_{i}}r_{i}^{2}\frac{\partial}{\partial r_{i}}-\frac{L_{i}^{2}}{r_{i}^{2}}
\end{equation}
where $L_{i}$ is the angular momentum operator.  As
\begin{equation}
P_{l}(cos(\theta))=\frac{4\pi}{(2l+1)}\sum_{m=-l}^{m=+l}Y_{lm}(\theta_{1},\phi_{1})Y_{lm}^{*}(\theta_{2},\phi_{2}),
\end{equation}
then
\begin{equation}
L_{i}^{2}P_{l}=l(l+1)P_{l}.
\end{equation}
Hence the kinetic terms are also $1$D integrals in $r_{1}$ and $r_{2}$ when $l'=l$, otherwise the integrals are zero.

The Coulomb interaction integral is more complicated.  First we expand the Coulomb interaction term using the Legendre polynomials
\begin{equation}
\frac{1}{|\bm{r}_{1}-\bm{r}_{2}|}=\sum_{k=0}^{\infty}\frac{(r_{<})^{k}}{(r_{>})^{k+1}}P_{k}(\cos(\theta)),
\end{equation}
where $r_{<}=\min\{r_{1},r_{2}\}$ and $r_{>}=\max\{r_{1},r_{2}\}$.

Next we use a result \cite{3LEGENDRE} for an integration of three Legendre polynomials
\begin{equation}
\int_{-1}^{1}P_{l'}(z)P_{k}(z)P_{l}(z)  dz= 2
\begin{pmatrix}l' & k & l \\ 0 & 0 & 0\end{pmatrix}^{2}.
\end{equation}

If $l'>k+l$ then the polynomial of degree $k+l$ can be formed from Legendre polynomials each of degree less than $l'$ hence the integral is zero, similarly for $k>l+l'$ and $l>k+l'$.  Here the $3j$ symbol (see for example Ref.~\cite{Messiah}) is zero for $l'+k+l=2p+1$ where $p$ is an integer. While for $l'+k+l=2p$ 
\begin{equation}
\begin{pmatrix} l' & k & l \\ 0 & 0 & 0\end{pmatrix}
=(-1)^{p}\sqrt{\Delta(l'kl)}\frac{p!}{(p-l')!(p-k)!(p-l)!},
\end{equation}

where
\begin{equation}
\Delta(abc)=\frac{(a+b-c)!(b+c-a)!(c+a-b)!}{(a+b+c+1)!}.
\end{equation}

Hence the integral for the Coulomb interaction when switching to spherical polar integration, including the normalisation, and integrating over angles is

\begin{eqnarray}
\nonumber\bra{\phi_{i'j'l'}} \frac{1}{|\bm{r}_{1}-\bm{r}_{2}|}\ket{\phi_{ijl}}=\\
\nonumber\sqrt{2l'+1}\sqrt{2l+1}\sum_{k=0}^{l+l'}\begin{pmatrix}l' & k & l \\ 0 & 0 & 0\end{pmatrix}^{2}\int_{0}^{\infty}dr_{1}R_{i'}(r_{1})R_{i}(r_{1})\\
\nonumber\big(\frac{1}{r_{1}^{k-1}}\int_{0}^{r_{1}}R_{j'}(r_{2})R_{j}(r_{2})r_{2}^{k+2} dr_{2}+\\
r_{1}^{k+2}\int_{r_{1}}^{\infty}R_{j'}(r_{2})R_{j}(r_{2})\frac{1}{r_{2}^{k-1}}dr_{2}  \big).
\end{eqnarray}

As the 3j symbol is zero for $k>l+l'$ then there are at most $l+l'$ double integrals, but the other constraints then produce a zero 3j symbol for many of these.  Hence efficiency is increased: we have a small sum of double integrals rather than a triple integral to compute.

We also consider the single particle KS equation (Eq.~\ref{eq:KS1}) where $v_{\text{ext}}(\bm{r})=-Z/r$ for helium-like atoms and $v_{\text{ext}}(\bm{r})=\omega r^{2}/2$ for Hooke's atom.
For approximating $v_{\text{xc}}$ we use $v^{\text{LDA}}_{\text{xc}}=v^{\text{LDA}}_{\text{x}}+v^{\text{LDA}}_{\text{c}}$.  We employ the fit of Perdew and Wang \cite{PERDEW92} (see Eq.~\ref{eq:PerdewWang}) for $v^{\text{LDA}}_{\text{c}}$ while the expression for $v^{\text{LDA}}_{\text{x}}$ is that of Wigner and Seitz \cite{WIGNER} (see Eq.~\ref{eq:WignerSeitz}).
We consider only spherically symmetric potentials so the KS equation reduces to a one dimensional problem. We solve this self consistently to obtain  $n^{\text{target}}$, the LDA approximation to the  ground state electron density.

 This `target' density is then used in the iterative scheme developed in Ref.~\cite{Coe2} which allows to find the external potential for the interacting many-body system of ground state density $n^{\text{target}}$, in this case the i-LDA system. This scheme is based on the relation
\begin{eqnarray}
v^{i+1}_{\text{ext}}(\bm{r_{1}})=\frac{1}{n_i(\bm{r_{1}})}|E_{i}|[n_{i}(\bm{r_{1}})-n^{\text{target}}(\bm{r_{1}})]
+v^{i}_{\text{ext}}(\bm{r_{1}}).
\label{eq:bothscheme}
\end{eqnarray}
 Here $E_{i}$ is the energy of the interacting system with external potential $\sum_j v^{i}_{\text{ext}}(\bm{r_{j}})$: this many-body system is solved at every step  by using the aforementioned method to give the many-body wavefunction from which we compute the density $n_{i}$.  We aid convergence for both test-bed systems by mixing $v^{i+1}_{\text{ext}}$ with $80\%$ of $v_{\text{ext}}^i$, and iterate until the error $\int d^3r\, |n_i(\bm{r})-n^{\text{target}}(\bm{r})|/\int d^3r\,n^{\text{target}}(\bm{r})$ has reached a desired level on the chosen integration range.

We note that the inversion scheme Eq.~\ref{eq:bothscheme} could also be used to derive the unknown potential related to a generic density $n^{\text{target}}$ obtained, e.g., from experimental measurements.

\section{Spatial entanglement calculation}
For the systems we are considering, the ground state wave-function can be factorised in orbital and spin part.
The spin part is a singlet which is maximally entangled and constant.  We therefore ignore the spin entanglement and investigate the spatial entanglement, i.e., the particle-particle entanglement stemming from the continuous spatial degrees of freedom.

We use the two-electron wavefunction to calculate the linear entropy of the reduced density matrix 
\begin{equation}
L=1-Tr\rho_{A}^{2}
\end{equation}
 as a measure of the spatial entanglement. Here $\rho_{A}=Tr_{B}\left|\psi \right\rangle \left \langle \psi \right|$ is found by tracing out the system density matrix over the degrees of freedom of one of the two subsystems.  The linear entropy was, perhaps, first put forward as a measure of purity of a quantum state by Zurek, Habib and Paz \cite{ZUREK93}.  The purity of the reduced density matrix can be shown to give an indication of the number and spread of the terms in the Schmidt decomposition of the state and therefore a measure of the entanglement.   Here, in the continuous case, we calculate the entanglement using
\begin{equation}
\rho_{\text{A}}(\bm{r_{1}},\bm{r_{2}})=\int \Psi^{*}(\bm{r_{1}},\bm{r_{3}})\Psi(\bm{r_{2}},\bm{r_{3}})\bm{dr_{3}},
\end{equation}
\begin{equation}
\rho^{2}_{\text{A}}(\bm{r_{1}},\bm{r_{2}})=\int \rho_{\text{A}}(\bm{r_{1}},\bm{r_{3}})\rho_{\text{A}}(\bm{r_{3}},\bm{r_{2}})\bm{dr_{3}},
\end{equation}
and
\begin{equation}
Tr\rho^{2}_{\text{A}}=\int \rho^{2}_{\text{A}}(\bm{r},\bm{r})\bm{dr}. 
\end{equation}

We will investigate the change in spatial entanglement with the strength of the confining potential, which is characterised by  $\omega$ or the nuclear charge $Z$ for Hooke's and helium-like atoms respectively.

\section{Helium-like atoms}

\begin{figure}[ht]\centering
  \includegraphics[width=.6\textwidth]{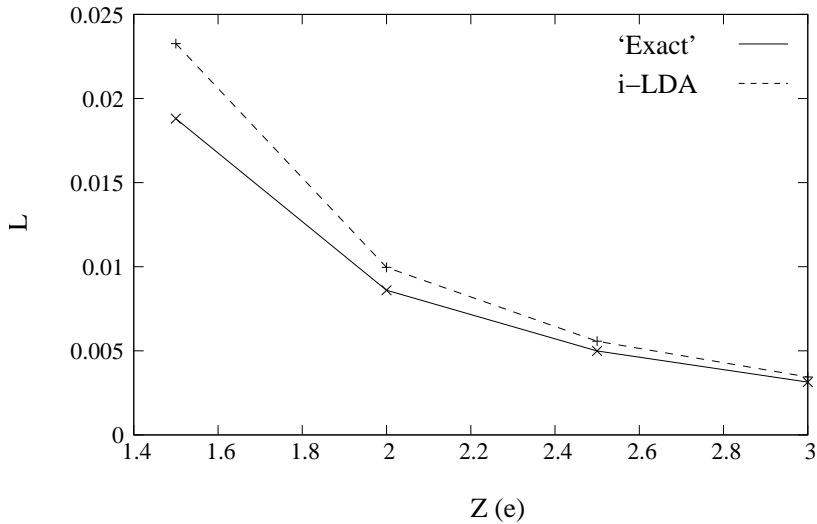}
  \caption{The linear entropy $L$ of the reduced density matrix versus the nuclear charge $Z$ as a measure of entanglement for helium-like atoms and the corresponding i-LDA systems.}\label{fig:NewHeliumEntanglement}
\end{figure}

For helium-like atoms the (numerically) exact and the i-LDA entanglement are depicted in Fig.~\ref{fig:NewHeliumEntanglement}.  There we see that the entanglement increases as the nuclear charge decreases which may be attributed to the increase in the ratio of the electron-electron interaction to the energy from the external potential. A similar situation was observed in the spatial entanglement of Hooke's atom \cite{Coe}.  We note that the spatial entanglement is low for the parameters considered, when compared with the theoretical maximum of $L=1$.  The `exact' and i-LDA entanglement are close and the gap between them increases as the weight of the particle-particle interactions increase.  This fits in with the LDA performing better when many-body interactions are weak: as the confining potential increases,
many-body interactions will become less important relative to it, until the system may be fairly well approximated by a non-interacting Hamiltonian thereby giving rise to a product state solution. 
We note that although helium-like atoms have a bound state until $Z\approx 0.911$ we were unable to numerically solve the Kohn-Sham equations with the LDA satisfactorily for $Z<1.5$.  We hypothesise that for the exact solution with $Z \lesssim 0.911$ a transition to a resonance state---as studied recently for two electrons in a spherically symmetric square well \cite{OSE3} and the spherical helium atom \cite{OSE2}---would be expected to occur.  This would result in a superposition of one electron constrained by the nucleus with the other unbound.  Such a wavefunction would be expected to have a spatial entanglement of $L=1/2$.

\section{Hooke's atom}
\begin{figure}[ht]\centering
  \includegraphics[width=.6\textwidth]{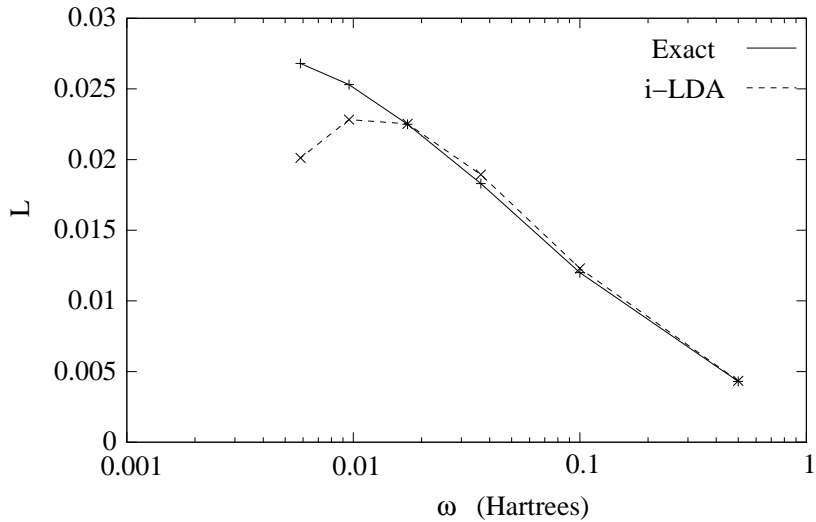}
  \caption{The linear entropy $L$ of the reduced density matrix  versus $\omega$ as a measure of entanglement for Hooke's atom  and the corresponding i-LDA system.}\label{fig:LDAentanglement}
\end{figure}

For Hooke's atom (Fig.~\ref{fig:LDAentanglement}) we see that the accuracy of the entanglement when using the LDA is high, but decreases a little as $\omega$ decreases and interactions, as quantified by the ratio of the electron-electron interaction to the energy from the external potential, become larger.  The value of the i-LDA entanglement for $\omega=0.00584$, lower than the value for $\omega=0.00957$, seems anomalous as does the change from an overestimate to an underestimate of the exact entanglement.  We note that $\omega=0.00584$ has the poorest match ($0.053\%$ error) between the i-LDA estimate and the actual LDA density of the $\omega$ displayed and is the smallest $\omega$ for which we could achieve a good match with a reasonable basis size of $8^{3}$ wavefunctions.  Therefore due to a seemingly high sensitivity of the entanglement to the wavefunction yet a relative insensitivity of the density to the wavefunction, we may have crossed a threshold so the small error in reproducing the LDA density is now not small enough to prevent a large error when calculating the LDA entanglement.  However we also note that $0.053\%$ error still represents a very good accuracy in  reproducing the LDA density, and that when employing an even larger basis the difference between old and new density estimates was just $0.089\%$. A basis of $8^{3}$ was also sufficient to calculate the exact entanglement to $3$ significant figures.  In addition we saw in previous work \cite{Coe} that the difference between exact and LDA energies exhibited a non-trivial relationship with $\omega$, and a perturbed LDA-based approach to approximate Hooke's atom many-body wavefunction also produces a decrease of the entanglement in a similar range of $\omega$ (see Fig.~\ref{fig:hookesiLDAandEA1L}).  These observations may imply that a non-trivial  relationship between $\omega$ and LDA-based estimates of the entanglement could be a real effect. We suggest that although the accuracy of the i-LDA entanglement for the two smallest values of $\omega$ may be questionable, the correct i-LDA entanglement for these values should not be too different and we are confident of the accuracy of the i-LDA entanglement for larger values of $\omega$.

\subsection{Comparison with other LDA-based entanglement estimates}
In previous work \cite{Coe} we devised different ways to derive an LDA-based approximation to the entanglement. 

The first is a combination of gradient descent and evolutionary algorithms to derive an approximation for the i-LDA wavefunction from which to calculate the entanglement. The main limitation of this approach is that the trial i-LDA wavefunction was based on a specific functional form. This variational-type approach included some parameters to be varied to improve performance. 

The second approach to derive an LDA-based approximation for the interacting many-body wavefunction is based on a perturbative scheme whose zero-order Hamiltonian includes the Kohn-Sham effective potential (and hence, through $v_{xc}$, partly the many-body interaction) with the LDA approximation for the exchange-correlation term. This approach improves as higher orders are taken into account, but a first-order wavefunction would still fail to reproduce the exact entanglement when many-body interactions become dominant.  

\begin{figure}[ht]\centering
  \includegraphics[width=.6\textwidth]{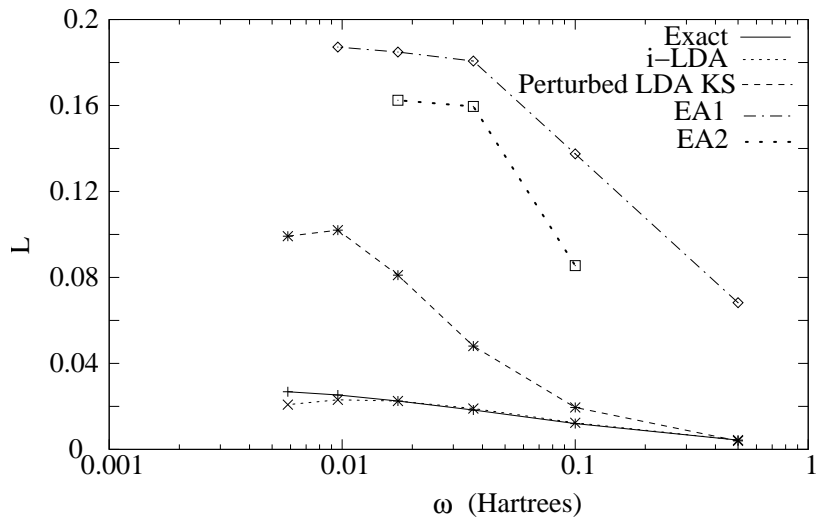}
  \caption{The linear entropy $L$ of the reduced density matrix as a measure of the entanglement obtained using the exact, i-LDA, first order perturbation of the LDA KS equations, and evolutionary algorithm (EA1 and EA2) many-body wavefunctions.}\label{fig:hookesiLDAandEA1L}
\end{figure}

We see in Fig.~\ref{fig:hookesiLDAandEA1L} that, in the intermediate and strong particle-particle interaction regime, the exact i-LDA entanglement for Hooke's atom reproduces the system exact entanglement much more accurately than any of the approximations mentioned above. Approximations labelled as EA1 and EA2 were achieved by using a combination of gradient descent and evolutionary algorithms~\cite{Coe}. In EA1 we sought to minimise the difference between the density arising from the trial wavefunction and the exact density and then, within the wavefunctions giving an accurate enough match, we choose the wavefunction corresponding to the lowest ground state energy.  The trial wavefunctions in this case are  restricted to wavefunctions which can be separated in centre of mass and relative motion components.
In EA2 we sought to minimise the difference between the density arising from the trial wavefunction and the exact density together with a minimisation of the energy of the trial wavefunction.  The trial wavefunctions in this case used a finite number of functions in terms of $r_{1}$, $r_{2}$, and $\theta$ from a general basis for two electrons in a spherically symmetric potential.  The gradient descent allows the basis coefficients for the wavefunction that gives a local minimum to be found while the evolutionary algorithm allows more of the parameter space to be explored so in principle, given a large enough basis and enough time, the combination should find the global minimum.  With our finite basis and limited run-time we see that, although this method gives the trend of the i-LDA entanglement, the approximate i-LDA entanglement from the evolutionary algorithm  is much higher than the actual i-LDA entanglement.  

As mentioned above, in~\cite{Coe} we also considered a perturbative approach based on the LDA-KS equations, whose related interacting wavefunction converges toward the exact wavefunction as higher orders are considered.  
To first order we would hope for a good approximation to the exact  entanglement that perhaps leans towards the i-LDA entanglement.  We see in Fig.~\ref{fig:hookesiLDAandEA1L} that the entanglement results from  first order becomes much greater than the i-LDA entanglement---but less so than the evolutionary algorithm---for stronger interactions, as quantified by the ratio of expectation of the interaction energy to the potential energy.  However, the perturbation method allows an easier and more efficient way to approximate the entanglement within an LDA-related scheme, whose first order  is accurate for low interactions;  the exact i-LDA entanglement, although more accurate, is inaccessible without repeated complicated calculations: the inversion scheme requires solving the many-body equations more times than would be needed to ascertain the exact entanglement.

\section{Position space information entropy}

We also investigate the position-space information entropy of the density
\begin{equation}
S_{n}=-\int n(\bm{r})\ln n(\bm{r})\bm{dr}.
\label{eqn:infentropy}
\end{equation}
as a proxy measure for the entanglement.  This has been used to study the entanglement of the Moshinsky atom \cite{AMOVILLI04} and may also be thought of as a zeroth-order approximation to the von Neumann entropy of the reduced density matrix, where the off-diagonal terms of the reduced density matrix are set to zero \cite{Coe}.  As the von Neumann entropy of the reduced density matrix may be used as a measure of the entanglement and has been seen to behave similarly to $L$ for Hooke's atom \cite{Coe},  it would be hoped that $S_{n}$ should capture some of the behaviour of the entanglement. 
\begin{figure}[ht]\centering
  \includegraphics[width=.6\textwidth]{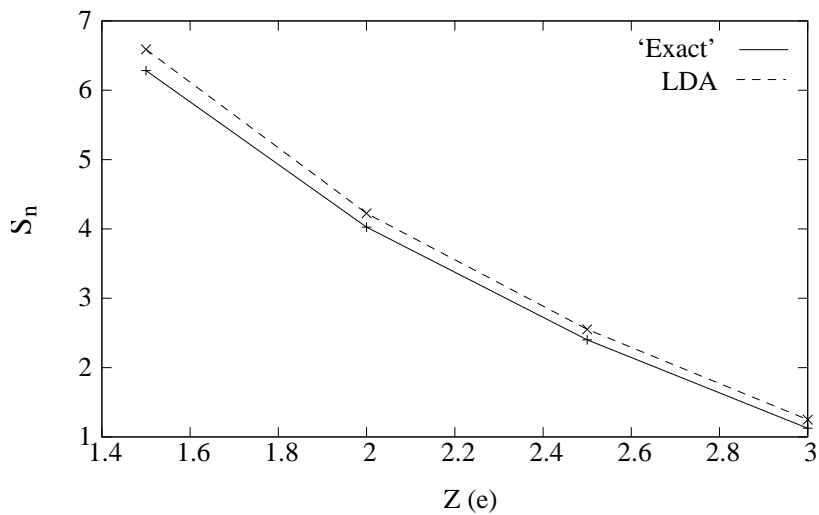}
  \caption{The position space information entropy $S_{n}$ for the `exact' and LDA densities for helium-like atoms.}\label{fig:HeliumInfoEntropy}
\end{figure}

We see in Fig.~\ref{fig:HeliumInfoEntropy} that $S_{n}$ captures the monotonically decreasing trend of the exact and i-LDA entanglement.  The $S_{n}$ for the LDA is greater than the exact $S_{n}$ but the difference between the two does not seem to increase as $Z$ decreases as is the case for the exact and i-LDA entanglement. The behaviour shown by $S_{n}$ is also more linear with increasing $Z$ than the entanglement.

\begin{figure}[ht]\centering
  \includegraphics[width=.6\textwidth]{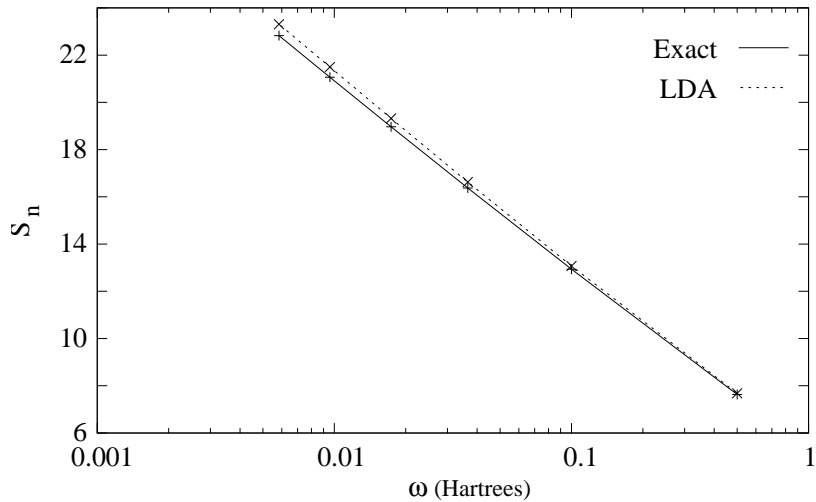}
  \caption{The position space information entropy $S_{n}$ for the `exact' and LDA densities for Hooke's atom.}\label{fig:HookesInfoEntropy}
\end{figure}

For Hooke's atom we previously noted that $S_{n}$ did not model the entanglement behaviour well over a much larger range of confining potentials \cite{Coe}\footnote{Label on $y$-axis in \cite{Coe}, Figs.~5 and 12, should read $S_{n}/\ln 2$.}.  We see in Fig.~\ref{fig:HookesInfoEntropy} that, over a smaller range of $\omega$, $S_{n}$ reproduces the general trend of the entanglement but $S_{n}$ from LDA is always greater than the exact $S_{n}$ which appears to not be the case for the entanglement (see Fig.~\ref{fig:LDAentanglement}).  In addition we see that $S_{n}$ is almost a straight line when plotted against $\omega$ on a logarithmic scale, this is in contrast to the entanglement which can be seen to increase less fast as $\omega$ decreases even for the small range of $\omega$ considered here.

\section{Conclusion}
We have calculated the spatial entanglement of the electrons in helium-like atoms corresponding to the exact system and to the use of the local-density approximation  within density-functional theory using the concept of the interacting LDA system.  The entanglement was seen to decrease with increasing nuclear charge $Z$.  The maximum entanglement found was only a fraction $\sim 2.5\%$ of the theoretical maximum.  The entanglement of the i-LDA system  was seen to reproduce the qualitative and quantitative behaviour of the exact entanglement with relatively good accuracy, although the discrepancy between the measures increased as the nuclear charge diminished.  We also compared the exact spatial entanglement of Hooke's atom to the entanglement of the related i-LDA system and saw that the LDA was a reasonably good approximation to the entanglement, but the accuracy decreased as the strength of the confining potential decreased and the i-LDA entanglement became non-monotonic at very small confining potential.  

We also saw that previously considered approximations to the exact or i-LDA entanglement were not so reliable when the particle-particle interaction became dominant, but noted that they captured the trend of the entanglement, were successful at strong confining potentials and were often much simpler computationally. In particular the accuracy of the perturbative approach based on the LDA-KS equations could be systematically improved by going beyond first order.

We considered the position space information entropy of the density, and show that it was able to efficiently give the general trend in the entanglement but not its detailed behaviour.

From our test-bed calculations on few-particle systems it would appear that the LDA may reproduce the entanglement behaviour well, even for systems with a highly inhomogeneous density and fairly strong interactions. More studies are necessary to ascertain if this conclusion holds for many-particle systems.

\section{Acknowledgement}

We acknowledge funding from EPSRC grant EP/F016719/1.

\providecommand{\newblock}{}


\begin{thebibliography}{10}
\expandafter\ifx\csname url\endcsname\relax
  \def\url#1{{\tt #1}}\fi
\expandafter\ifx\csname urlprefix\endcsname\relax\def\urlprefix{URL }\fi
\providecommand{\eprint}[2][]{\url{#2}}

\bibitem{KS}
Kohn W and Sham L~J 1965 {\em Phys. Rev.\/} {\bf 140} 1133

\bibitem{HK}
Hohenberg P and Kohn W 1964 {\em Phys. Rev.\/} {\bf 136} B864

\bibitem{Coe}
Coe J~P, Sudbery A and D'Amico I 2008 {\em Phys. Rev. B\/} {\bf 77} 205122

\bibitem{Zalka98}
Zalka C 1998 {\em Fortschr. Phys.\/} {\bf 46} 877

\bibitem{NIELSEN}
Nielsen M~A and Chuang I~L 2000 {\em Quantum Computation and Quantum
  Information\/} (Cambridge University Press)

\bibitem{DFTREVIEW}
Capelle K 2006 {\em Brazilian Journal of Physics\/} {\bf 36} 1318

\bibitem{WIGNER}
Wigner E~P and Seitz F 1933 {\em Phys. Rev.\/} {\bf 43} 804

\bibitem{PERDEW92}
Perdew J~P and Wang Y 1992 {\em Phys. Rev. B\/} {\bf 45} 13244

\bibitem{CEPALDER80}
Ceperley D~M and Alder B~J 1980 {\em Phys. Rev. Lett.\/} {\bf 45} 566--569

\bibitem{Coe2}
Coe J~P, Capelle K and D'Amico I 2009 {\em Phys. Rev. A\/} {\bf 79} 032504

\bibitem{BAKER}
Baker J~D, Freund D~E, Hill R~N and Morgan\:III J~D 1990 {\em Phys. Rev. A\/}
  {\bf 41} 1247

\bibitem{3LEGENDRE}
Mavromatis H~A and Alassar R~S 1999 {\em Applied Mathematics Letters\/} {\bf
  42} 101--105

\bibitem{Messiah}
Messiah A 1999 {\em Quantum Mechanics\/} (Dover Publications, Inc.)

\bibitem{ZUREK93}
Zurek W~H, Habib S and Paz J~P 1993 {\em Phys. Rev. Lett.\/} {\bf 70} 1187

\bibitem{OSE3}
Ferr\'{o}n A, Osenda O and Serra P 2009 {\em Phys. Rev. A\/} {\bf 79} 032509

\bibitem{OSE2}
Osenda O and Serra P 2007 {\em Phys. Rev. A\/} {\bf 75} 042331

\bibitem{AMOVILLI04}
Amovilli C and March N~H 2004 {\em Phys. Rev. A\/} {\bf 69} 054302

\end{thebibliography}
\end{document}